\documentclass[10pt,twocolumn,aps,pra,showpacs,amsmath,amssymb,superscriptaddress]{revtex4-1}
\usepackage{times}
\usepackage{color}
\usepackage{graphicx}
\usepackage{bm}
\usepackage{amsmath}
\usepackage{amsfonts}
\usepackage{amsthm}
\usepackage{amssymb}
\usepackage{mathrsfs}
\begin{document}
\title{Tight Bounds for the Entanglement of Formation of Gaussian States }
\author{Fernando Nicacio}
\email{fernando.nicacio@ufabc.edu.br  } 
\affiliation{ 
Centro de Ci\^encias Naturais e Humanas, 
Universidade Federal do ABC,   09210-170  , Santo Andr\'e, S\~ao Paulo, Brazil}
\author{Marcos C. de Oliveira}
\affiliation{
Instituto de F\'{\i}sica Gleb Wataghin, 
Universidade Estadual de Campinas, 
13083-859, Campinas, S\~{a}o Paulo, Brazil}
\date{\today}
%
\begin{abstract}
We establish tight upper and  lower bounds for the Entanglement of Formation of an
arbitrary two-mode Gaussian state employing the necessary properties of Gaussian channels.
Both bounds are strictly given by the Entanglement of 
Formation of symmetric Gaussian states, which are simply constructed from the 
reduced states obtained by partial trace of the original one. 
\end{abstract}
\maketitle
\section{Introduction}\label{introd}
A considerable effort has been devoted to the characterization of correlations contained in 
quantum states, or how much information two parts of the same system can share. 
The nature of these correlations can be classical or genuinely quantum, the last one being
characterized by the presence of some sort of entanglement \cite{RevModPhys.81.865}. 
For pure bipartite states (states solely quantum correlated)
this question was solved a long time ago: every measure of entanglement is completely 
equivalent to the von Neumann entropy of the reduced state of the 
bipartite system --- It quantifies how much shared information 
the global system loses after a partial trace.
On the other hand, when both kinds of correlations are present, {\it i.e.}, 
when dealing with mixed states, 
it is impossible to know which kind of correlations 
were lost after the partial trace. The best we can do is to minimize over all 
possible quantum correlated state decompositions present in this mixed one ---
the process called as the convex roof of a measure.  
The convex roof of the von Neumann entropy is what we call 
Entanglement of Formation (EoF). 
Among all measures of entanglement the EoF plays a fundamental role: 
based on the principle that entanglement cannot increase under 
local operations it was shown that this measure is a lower bound for 
all suitable measures of entanglement \cite{PhysRevLett.84.2014}. 
Theoretically, the convex roof extension of a measure is very well defined, 
but in practice it is hard to solve. 
Only for a small class of states presenting special symmetries is it possible to express the EoF analytically \cite{RevModPhys.81.865}. 

Gaussian states (GS) are remarkable states in physics, and in quantum information theory 
they are the natural candidates to implement quantum computation with continuous 
variable states \cite{PhysRevLett.82.1784}. This argument is sufficient to understand 
the collective effort of the community to search for an analytical expression for GS EoF. 
The first step in that direction \cite{PhysRevLett.91.107901} considered symmetric Gaussian states (SGS), 
defined as states where both reduced partitions have equal purity or equal von Neumann entropies.  
%
Subsequently a definition of another 
convex roof extension --- the Gaussian Entanglement of Formation (GeoF)  appeared \cite{PhysRevA.69.052320}. 
There the minimization procedure is taken over a restricted set  --- the set of 
pure GS --- and therefore is equal to the EoF when the state is symmetric. 
However no analytical expression was given: the process relies on a minimization of a 
polynomial function. Another important conceptual step was presented in \cite{PhysRevA.69.012307}, 
where the authors found two distinct lower bounds to the EoF of GS
and showed the importance of knowing at least analytical bounds for the EoF. 
More recently, the work \cite{PhysRevLett.101.220403} 
shows that the set used in the numerical 
minimization procedure to calculate the GoeF is indeed the correct one to find 
the EoF for a GS. 

In this paper we show a new way to determine generic tight bounds for the EoF
of an arbitrary GS. We use the very known concept of classical Gaussian channels \cite{EisertWolf2005}
together with the desired convexity property of generic measures of entanglement. 
This paper is organized as follows: In Sec. II we define the set of 
GS and present some necessary concepts and quantities involved with 
the EoF calculation, whose properties are presented in Sec. III. 
In Sec. IV  we review the definition of Gaussian channels 
and in Secs. V and VI we present our central results on the 
derivation of the limits to the EoF. 
Finally in Sec VII we present our conclusions and perspectives.

\section{Gaussian States}
The covariance matrix (CM) of a genuine two-mode bipartite GS  
$\hat \rho_{\!{\mathbf A} {\!\mathbf B}}$ is defined by 
\begin{equation}                                                                         \label{gauss}
\mathbf V_{\!\!{\mathbf A} {\!\mathbf B}} \equiv 
\left( 
\begin{array}{lc}
\mathbf A & \mathbf C \\
\mathbf C^\top & \mathbf B
\end{array} \right),    
\end{equation}
where $\mathbf A,\mathbf B,\mathbf C$ are $2 \times 2$ block matrices,
with $ \mathbf B \ge \mathbf A \ge 0$, without loss of generality. 
As a manifestation of the Heisenberg uncertainty principle, 
this CM must regard the following (positivity semidefiniteness)
inequality 
\begin{equation}                                                                         \label{heis}
\mathbf V_{\!\!{\mathbf A} {\!\mathbf B}} + i \mathsf J \ge 0 
\,\,\, \textrm{where} \,\,\, 
\mathsf J = \left( \begin{smallmatrix}
                   0  & 1 \\
                 - 1  & 0 
                   \end{smallmatrix} \right) \oplus 
                   \left( \begin{smallmatrix}
                   0  & 1 \\
                 - 1  & 0 
                   \end{smallmatrix} \right). 
\end{equation}
The generalization of this inequality for many-modes is trivial 
and only enhances the dimensions of the CM and $\mathsf J$.  

Using unitary local operations (which do not change the degree of entanglement)
we can reduce the above state to the so called standard form \cite{PhysRevLett.84.2726}: 
\begin{equation}                                                                          \label{locgauss}
\mathbf A  \mapsto a \,\mathsf{I}_2,  \,\,\,                                
\mathbf B  \mapsto b \, \mathsf{I}_2   \,\,\, {\rm and} \,\,\, 
\mathbf C  \mapsto {\rm Diag}\left( c_1 , c_2 \right),  
\end{equation}
where $\mathsf{I}_2$ is the two dimensional identity matrix,  
$a,b \ge 1$  and, for simplicity, $c_1 \ge |c_2|, \,\, c_2 < 0$.                                                               
We also define the local symplectic invariants: 
\begin{eqnarray}                                                                          \label{inv}
I_1 & \equiv  & \det \mathbf A = a^2, \,\,\,  
I_2 \equiv \det \mathbf B = b^2,  \,\,\ 
I_3 \equiv  \det \mathbf C = c_1 c_2, \nonumber \\  
I_4 & \equiv & {\rm Tr} \left( 
               \bf A \mathsf J \bf C \mathsf J \bf B \mathsf J C^\top \mathsf J 
               \right) = ab(c_1^2 + c_2^2).    
\end{eqnarray}
Using the above definitions we are able to calculate the 
symplectic eigenvalues (SE) of the CM in (\ref{gauss}), 
as in \cite{0953-4075-37-2-L02}:  
\begin{equation}                                                                          \label{simpeig}
\mu_{\pm} = \sqrt{ \tfrac{I_1 + I_2}{2} +  I_3   \pm 
                    \sqrt{ \left( \tfrac{ I_1 - I_2 }{2} \right)^2 
                    + (I_1 + I_2) I_3 +  I_4 }  }.
\end{equation}
We could also arrange them as a diagonal matrix, 
$\Lambda_{\mathbf V_{\!\!{\mathbf A} {\!\mathbf B}}} \equiv 
\mu_-\, \mathsf{I}_2 \oplus \mu_+ \, \mathsf{I}_2$,  
which we call symplectic spectrum.
The imposition of (\ref{heis}) guaranties that a genuine physical 
state must obey $\mu_{+} \ge \mu_{-} \ge 1$. 

\ When the CM (\ref{gauss}) undergoes a partial transposition 
transformation, represented by the diagonal matrix 
$\mathbf T_{\!B} := \mathsf I_2 \oplus \hat{\sigma}_z $, 
where $\hat{\sigma}_z$ is the third Pauli matrix, 
it becomes 
$\widetilde{\mathbf V}_{\!\!{\mathbf A} {\!\mathbf B}} \equiv 
\mathbf T_{\!B} {\mathbf V}_{\!\!{\mathbf A} {\!\mathbf B}}\mathbf T_{\!B}^\top$.  
The net effect of this transposition is to change the signal of $c_2$ in (\ref{inv}), 
and the symplectic spectrum of $\widetilde{\mathbf V}_{\!\!{\mathbf A} {\!\mathbf B}}$ 
is simply obtained from (\ref{simpeig}) by the substitution $I_3 \mapsto - I_3$:
\begin{equation}                                                                          \label{simpeitp}
\widetilde{ \mu }_{\pm} = \sqrt{ \tfrac{I_1 + I_2}{2} -  I_3   \pm 
                    \sqrt{ \left( \tfrac{ I_1 - I_2 }{2} \right)^2 
                    - (I_1 + I_2) I_3 +  I_4 }  }.
\end{equation}
Applying the Peres-Horodecki separability criteria \cite{PhysRevLett.84.2726} 
to the CM (\ref{gauss})
a bipartite GS is entangled iff $\widetilde{ \mu }_{-} < 1$.   

\ Let us define the CM of a SGS 
$\hat \rho_{{\mathbf M} {\!\mathbf M}}$ as 
\begin{equation}                                                                         \label{simgauss}
\mathbf V_{\!\!{\mathbf M} {\!\mathbf M}} \equiv 
\left( 
\begin{array}{lc}
\mathbf M & \mathbf C \\
\mathbf C^\top & \mathbf M
\end{array} \right),    
\end{equation}
{\it i.e.}, Eq.~(\ref{gauss}) with 
${\mathbf A} = {\mathbf B} = {\mathbf M}$ which under a local transformation 
$\mathbf M \mapsto m \, \mathsf{I}_2$.
Its SE and the SE of
its partial transposition are obtained from (\ref{simpeig}) and (\ref{simpeitp}) 
and are, respectively, given by 
\begin{equation}                                                                          \label{simsimpeig}
\!\!\!\!\nu_{\pm} = \sqrt{I_1 +  I_3   \pm 
            \sqrt{ I_4 + 2 I_1 I_3 }  } = \sqrt{ (m \pm c_1) (m \pm c_2) }  
\end{equation}
and
\begin{equation}                                                                          \label{simsimpeitp}
\!\!\!\!\!\widetilde{ \nu }_{\pm} = \sqrt{I_1 -  I_3   \pm 
            \sqrt{ I_4 - 2 I_1 I_3 }  } = \sqrt{ (m \pm c_1) (m \mp c_2) }. 
\end{equation}
As we will see in the next section the EoF for 
SGS is a monotonically decreasing function whose 
argument is the smaller eigenvalue in 
(\ref{simsimpeitp}).

\section{Entanglement of Formation}
The EoF for a mixed state 
$\hat \rho = \sum_i p_i | \psi_i \rangle \! \langle \psi_i |$ 
is constructed as the convex roof of the von Neumann entropy $S$ for pure states: 
\begin{equation}                                                                         \label{eof}
{\rm EoF}( \hat \rho ) = \inf_{\{ p_i, \psi_i \} } \sum_{i} p_i S(\psi_i),  
\end{equation}
the set $\{ p_i, \psi_i \}$ indicates that the minimization runs over all physically 
possible decompositions of $\hat \rho$. 

\ Among all properties of the EoF defined above 
two of them will be very important for us: 
{\it locality } and {\it convexity} \cite{RevModPhys.81.865,PhysRevLett.84.2014}. 
The {\it locality} states that the action of a local operation 
cannot increase the EoF. Furthermore, the EoF does not change under 
unitary local operations, which may be summarized as: 
if $\hat{U}_L$ is a local unitary operator, then 
\begin{equation}                                                                         \label{eof1}
{\rm EoF} \left( \hat \rho \right) =  
{\rm EoF} \left(\hat U_L \, \hat \rho \, \hat U_L^{\dag} \right). 
\end{equation}
Now, let us define a set of $N$ real numbers $ 0 \le \alpha_i \le 1 \, \forall i$,  
such that $\sum_{i = 1}^{N} \alpha_i = 1$ so that one can construct
a convex decomposition of $\hat \rho$ into a set of density matrices $\hat \rho_i$.    
The {\it convexity} of the EoF implies that
\begin{equation}                                                                         \label{eof3}
{\rm EoF} \left( \sum_{i = 1}^N  \alpha_i \hat \rho_i \right) \le 
 \sum_{i = 1}^N  \alpha_i \, {\rm EoF} \left( \hat \rho_i \right).
\end{equation}

Working directly on formula (\ref{eof}), using the above two properties
and the von Neumannn entropy of squeezed states, 
the authors in Refs.~\cite{PhysRevLett.91.107901} could obtain an analytical formula 
for the EoF of any two mode SGS as
\begin{equation}                                                                         \label{simeof}
{\rm EoF}( \hat \rho_{{\mathbf M} {\!\mathbf M}} ) = 
f \left( \tilde{ \nu }_{-} \right), 
\end{equation}
where $ \tilde{ \nu }_{-}$ is the symplectic eigenvalue of the partially 
transposed CM $\widetilde{\mathbf V}_{\!\!{\mathbf M} {\!\mathbf M}}$ and the 
monotonically decreasing function $f$ is defined as 
$f(x) = c_+(x) {\rm ln}(c_+(x)) - c_-(x){\rm ln}(c_-(x))$ with
$c_{\pm}(x)=(x^{-1/2}\pm x^{1/2})^2/4$.
An attempt to generalize  (\ref{simeof}) for non-SGS with CM 
${\mathbf V}_{\!\!{\mathbf A} {\!\mathbf B}}$ is given by the adoption of the 
function $f$ of (\ref{simeof}) with $ \tilde\mu_{-}$ defined in (\ref{simpeig}) 
as the argument, we call this quantity EeoF:  
\begin{equation}                                                                         \label{eeof}
{\rm EeoF}( \hat \rho_{ {\! \mathbf A} { \! \mathbf B } } ) = 
f \left( \tilde{ \mu }_{-} \right). 
\end{equation}
In Refs.~\cite{1751-8121-40-28-S01,0953-4075-37-2-L02}
the authors conjecture that this should be the expression of the true EoF for GS,
but here we will argue in the next sections that this quantity 
can be considered an estimation for the EoF.

\ It is possible to define a bona fide measure of entanglement 
even when the states $| \psi_i \rangle$ into the decomposition in (\ref{eof}) 
are taken to be Gaussian \cite{PhysRevA.69.052320}. 
This measure is known as Gaussian Entanglement of Formation (GeoF) and 
there isn't an analytical expression for it. Indeed, it should be calculated by
a minimization of a polynomial function whose coefficients are cumbersome functions of 
the entries of the matrix ${\mathbf V}_{\!\!{\mathbf A} {\!\mathbf B}}$ and   
are explicitly written in Ref.~\cite{1751-8121-40-28-S01}. 
As a matter of fact, in Ref.~\cite{PhysRevLett.101.220403} the authors show
that the GeoF is the EoF for Gaussian states.

\section{Gaussian Channels}
The Gaussian channels (GC) considered here are trace preserving and completely 
positive maps acting on density operators, preserving also the Gaussian character 
of a state of this kind \cite{EisertWolf2005,1367-2630-10-8-083030}. 
We will only concern ourselves with the 
classical noise channel (CNC), whose action on a density 
operator can be written as a convolution of the density operator with 
a Gaussian \cite{CavesWodf2004}, {\it i.e.}, 
\begin{equation}                                                                         \label{bosonico}
\hat \rho = \frac{1}{(2\pi)^2} \int^{+ \infty}_{-\infty} 
\frac{ {\rm e}^{-\frac{1}{\hbar} \xi \cdot \Delta^{-1} \xi} } 
     { \sqrt{\rm Det \Delta} } \,\, 
\hat T_\xi \hat\rho_0 \hat T_\xi^\dag \,\,\, d^{4} \xi.                                               
\end{equation}
The operators $\hat T_\xi$ are the 
Weyl displacement operators \cite{deAlmeida1998265}.
The vector $\xi \in \mathbb R^4$ and $\Delta$ must be a positive 
semidefinite matrix, $\Delta \ge 0$. 
Physically, the noise channel may be implemented as the interaction 
of the system with a thermal bath at high temperature. 

\ Concerning the CM of the states involved in (\ref{bosonico}), 
it is easy to show that if $\hat \rho_0$, which not necessarily Gaussian, 
has a CM $\mathbf V_0$, the state $\hat \rho$ will have the CM 
\begin{equation}
\mathbf{ V } = \mathbf V_0 + \Delta. 
\end{equation}
Since the sum of positive semidefinite matrix is positive semidefinite, 
if $\mathbf V_0$ obeys (\ref{heis}), then $\mathbf V$ also will. 
From Eq.~(\ref{bosonico}) one can see that it is a convex sum of operators
once
\begin{equation}
\frac{{\rm e}^{-\frac{1}{2} \xi \cdot \Delta^{-1} \xi}} {\sqrt{\rm Det \Delta}} 
\ge 0 \,\,\, {\rm and} \,\,\,  
\int^{+ \infty}_{-\infty} 
\frac{{\rm e}^{-\frac{1}{2} \xi \cdot \Delta^{-1} \xi}} 
     {(2\pi)^2  \sqrt{\rm Det \Delta}} \,\, 
 d^{4} \xi = 1.                                                
\end{equation}
Now one can use Eq.~(\ref{eof3}) for the convex sum in (\ref{bosonico}) 
and the locality of the Weyl operator, Eq.~(\ref{eof1}), to show that
\begin{equation}                                                                         \label{eof4}
{\rm EoF} \left( \hat \rho \right) \le
{\rm EoF} \left(  \hat \rho_0  \right).                                                  
\end{equation} 
In such a way, one can conclude that 
\begin{equation}                                                                         \label{convLim}
\mathbf V = \mathbf V_{0} + \Delta 
\Longrightarrow  
{\rm EoF} \left( \hat \rho \right) \le 
{\rm EoF} \left( \hat \rho_{0}  \right), \,\,\, \forall \Delta \ge 0. 
\end{equation}
This equation is the principal statement of the present work, 
it will be useful for finding lower and upper bounds for the EoF of a general GS   
and it has a clear physical interpretation: as the channel adds noise to the 
system, there is no strangeness if the quantum correlations diminish.

\section{Bounds for EoF}
Let us consider two SGS, 
$\hat \rho_{{\mathbf N} {\!\mathbf N}}$,  
$\hat \rho_{{\mathbf M} {\!\mathbf M}}$, whose CV are of the form (\ref{simgauss})
and a non symmetric one, 
$\hat \rho_{{\!\mathbf A} {\!\mathbf B}}$,
whose CV is of the form given in (\ref{gauss}). 
Mind that in our notation, 
all of the above states have a CM with the same block matrix $\mathbf C$. 
Suppose the following order to the matrices: 
\begin{equation}                                                                         \label{ordmat}
\mathbf N \ge \mathbf B \ge \mathbf A \ge \mathbf M.
\end{equation}
Now we are able to find bounds for the EoF of a generic GS 
$\hat \rho_{\!\mathbf A\!\mathbf B} $. 
First, let us define two noise matrices
$\Delta_1 \equiv ( \mathbf N - \mathbf A ) \oplus ( \mathbf N - \mathbf B )$ 
and 
$
\Delta_2  \equiv ( \mathbf A - \mathbf M ) \oplus ( \mathbf B - \mathbf M ) $, 
both are positive semidefinite regarding the ordering imposed in (\ref{ordmat}). 
It is easy to see that 
\begin{equation}                                                                       \label{ord1a}                                                                         
\mathbf V_{\!\!{\mathbf N} {\!\mathbf N}}  =   
\mathbf V_{\!\!{\mathbf A} {\!\mathbf B}} + \Delta_1 
\,\,\,\,\,\, \text{and} \,\,\,\,\,\,
\mathbf V_{\!\!{\mathbf A} {\!\mathbf B}} = 
\mathbf V_{\!\!{\mathbf M} {\!\mathbf M}} + \Delta_2,   
\end{equation}                                                                   
therefore using the statement in (\ref{convLim}), 
we can sort the EoF for the states as
\begin{equation}                                                                         \label{ord1}
  {\rm EoF} \left( \hat \rho_{  \mathbf N\!\mathbf N}  \right) \le 
  {\rm EoF} \left( \hat \rho_{\!\mathbf A\!\mathbf B}  \right) \le
  {\rm EoF} \left( \hat \rho_{  \mathbf M\!\mathbf M}  \right).     
\end{equation}
The advantage of the limiting bounds can be seen by the fact that they are 
the EoF of SGS and can easily be calculated by (\ref{simeof}).
Note that the Gaussian channels described by the noise matrices in (\ref{ord1a}) 
are non unitary operations but Gaussian and local (GLOCC). 

As a matter of fact, until now we needed to assume that 
$\hat \rho_{{\mathbf M} {\!\mathbf M}}$ in (\ref{ord1}) represents a
genuine physical state in the sense of Eq.~(\ref{heis}). In view
of the sum of positive semidefinite matrices, 
Eq.~(\ref{ord1a}) implies  
\begin{equation}                                                                         \label{cond}
\mathbf V_{\!\!{\mathbf N} {\!\mathbf N}} + i \mathsf J \ge 
\mathbf V_{\!\!{\mathbf A} {\!\mathbf B}} + i \mathsf J \ge 
\mathbf V_{\!\!{\mathbf M} {\!\mathbf M}} + i \mathsf J \ge 0, 
\end{equation}
which means that the physicality imposed on the lower matrix 
guarantees the physicality for all the others.

\begin{figure}[htbp!] 
\includegraphics[width=8.5cm,trim = 0 50 0 40]{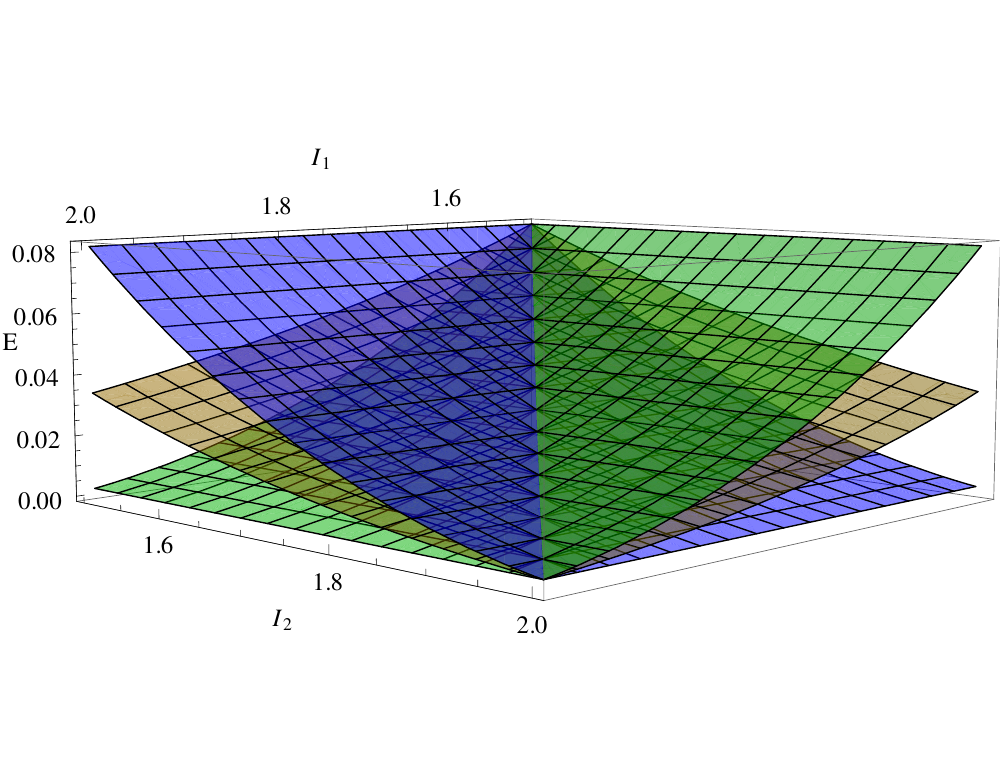} 
\caption{ 
GeoF function (orange) bounded by the EoF of symmetric states 
(blue and green) as a function of the local symplectic invariants 
$I_1$ and $I_2$. The other invariants are chosen to guarantee 
the existence and entanglement of the states: 
$I_3 = -0.2 $ and $I_4 = 2|I_3|\sqrt{I_1I_2}$. }            \label{fig1} 
\end{figure}

As a corollary of the result (\ref{ord1}), the EoF of a non symmetric 
Gaussian state GS with CV (\ref{gauss}) has two 
{\it natural bounds} 
\begin{equation}                                                                         \label{ord3}
  {\rm EoF}(\hat \rho_{{\!\mathbf B} {\!\mathbf B}}) \le 
  {\rm EoF} \left( \hat \rho_{\!\mathbf A\!\mathbf B}  \right) \le
  {\rm EoF}(\hat \rho_{{\mathbf A} {\!\mathbf A}}),     
\end{equation}
since 
\begin{eqnarray}                                                   
\mathbf V_{\!\!{\mathbf A} {\!\mathbf B}} &=& 
\mathbf V_{\!\!{\mathbf A} {\!\mathbf A}} + 
{\bf 0}_2 \oplus ({\mathbf B} - {\mathbf A})       \,\,\, \text{and} \,\,\,              \nonumber \\
\mathbf V_{\!\!{\mathbf B} {\!\mathbf B}} &=& 
\mathbf V_{\!\!{\mathbf A} {\!\mathbf B}} + 
({\mathbf B} - {\mathbf A}) \oplus {\bf 0}_2,
\end{eqnarray}
where ${\bf 0}_2$ is the $2\times2$ null matrix.
In figure \ref{fig1}, one can see the GeoF for a non symmetric
Gaussian state $\hat \rho_{\! \mathbf A \!\mathbf B}$, 
calculated by the recipe of \cite{PhysRevA.69.052320, 1751-8121-40-28-S01}
(remember that following \cite{PhysRevLett.101.220403}, 
this GeoF must be the true EoF for GS) bounded by the EoF of SGs 
$\hat \rho_{\! \mathbf A \!\mathbf A}$ and 
$\hat \rho_{   \mathbf B \!\mathbf B} $. 

\ Comparing the EPR-uncertainties \cite{PhysRevLett.91.107901} of mixed GSs 
and of squeezed states, the authors in \cite{PhysRevA.69.012307} 
obtained the EoF for a SGS 
$\hat \sigma$ whose CM is like (\ref{simgauss}) 
with $\mathbf M = (\mathbf A + \mathbf B)/2$ 
as a lower bound for the EoF of the general GS (\ref{gauss}):  
${\rm EoF} \left( \hat \sigma  \right) \le
 {\rm EoF} \left( \hat \rho_{\! \mathbf A \!\mathbf B}  \right)$. 
It is impossible to deduce this bound using (\ref{ord1}) 
since we can not construct $\Delta_1$ and $\Delta_2$ 
preserving positive semidefiniteness. 
%
%
However, 
$\mathbf A \le \mathbf M = (\mathbf A + \mathbf B)/2 \le \mathbf B$;
then comparing the CM of the states using (\ref{convLim}) with 
$\hat \rho = \hat \rho_{   \mathbf B \!\mathbf B}$, 
$\hat \rho_{0} = \hat \sigma$ 
and 
$ \Delta =  (\mathbf B - \mathbf M) \oplus (\mathbf B - \mathbf M) $, 
one can establish its value 
on the hierarchy of (\ref{ord3}) as 
\begin{equation}                                                                         \label{ord2}
    {\rm EoF} \left( \hat \rho_{   \mathbf B \!\mathbf B}  \right) \le 
    {\rm EoF} \left( \hat \sigma  \right) \le
    {\rm EoF} \left( \hat \rho_{\! \mathbf A \!\mathbf B}  \right) \le 
    {\rm EoF} \left( \hat \rho_{\! \mathbf A \!\mathbf A}  \right).     \!
\end{equation}
The closer lower bound given above is always a 
physical state \cite{PhysRevA.69.012307} which is the 
best lower bound allowed by our method, {\it i.e.}, 
any attempt to find a SGS with EoF closer to 
${\rm EoF} \left( \hat \rho_{\! \mathbf A \!\mathbf B}  \right)$ and 
smaller than $ {\rm EoF} \left( \hat \sigma  \right) $
fails to find a positive semidefinite $\Delta_1$ in (\ref{ord1a}).

\ Furthermore, given an arbitrary Gaussian state, 
some available relation between the local 
covariance matrices can be used to determine other bounds for the EoF 
of the original state, {\it e.g.}, suppose 
${\mathbf B} - {\mathbf A} \le {\mathbf A} $,
then the symmetric state 
$\hat \rho_{  \mathbf M\!\mathbf M } $ with 
$\mathbf M = {\mathbf B} - {\mathbf A} $ constitutes an upper bound to the 
state $\hat \rho_{\! \mathbf A \!\mathbf B}$.
As a final comment, 
nothing prevents the nonphysicality (even the nonpositivity) 
of the operator $\hat \rho_{\! \mathbf A \!\mathbf A}$ in (\ref{ord3}) 
when constructed from (\ref{gauss}). 
Remembering that in our protocol all the matrices have 
the same correlation matrix $\mathbf C$, see Eqs.~(\ref{gauss}) and (\ref{simgauss}), 
one way to detour this undesired behavior is to search for
another SGS described by a CM $\mathbf V'$ with a different correlation matrix 
but with $\Delta' = \mathbf V_{\!\!{\mathbf A} {\!\mathbf B}} - \mathbf V' \ge 0$. 
%

\section{Estimation for EoF}
Actually, we can derive a more general and mathematical precise procedure 
independent of Gaussian channels and physical states to determine 
an estimation for the EoF.
This criterion for the EoF functions is a 
direct consequence of the Williamson theorem \cite{deGosson2009131}: 
considering two positive semidefinite 
matrices, $\mathbf H_1 \ge \mathbf H_2$, 
their symplectic spectrum must be 
sorted as $\Lambda_{\mathbf H_{1}} \ge \Lambda_{\mathbf H_{2}}$.   
Assuming $f$ as a monotonically decreasing function, 
like the function defined below Eq.~(\ref{simeof}),  
one can see that 
\begin{equation}                                                                         \label{willianson}
{\mathbf H_1}  \ge 
{\mathbf H_2} \Longrightarrow
\tilde{\mathbf H}_1  \ge 
\tilde{\mathbf H}_2 \Longrightarrow 
f(\tilde{\nu}^{-}_2)  \ge f(\tilde{\nu}^{-}_1), 
\end{equation}
where 
$\tilde{\mathbf H}_i = \mathbf T_{\!B} \mathbf H_i \mathbf T_{\!B}$ 
is the partial transposition of the matrix ${\mathbf H_i}$ 
already defined and $\tilde{\nu}^{-}_i$ is the smaller 
symplectic eigenvalue of the matrix $\tilde{\mathbf H}_i$.  

The statement of Eq.~(\ref{willianson}) is sufficient to prove that the function 
${\rm EeoF}(\hat \rho_{\! \mathbf A \!\mathbf B} )$ defined in (\ref{eeof}) 
is also bounded exactly as ${\rm EoF}(\hat \rho_{\! \mathbf A \!\mathbf B} )$ 
in (\ref{ord2}) by the EoF of the same symmetric states. 
To see this let us take a look at the situation in Eq.~(\ref{cond}), 
\begin{equation}                                                                         \label{partorder}   
\mathbf V_{\!\!{\mathbf N} {\!\mathbf N}} \ge 
\mathbf V_{\!\!{\mathbf A} {\!\mathbf B}} \ge 
\mathbf V_{\!\!{\mathbf M} {\!\mathbf M}} 
\Longrightarrow
\widetilde{\mathbf V}_{\!\!{\mathbf N} {\!\mathbf N}} \ge 
\widetilde{\mathbf V}_{\!\!{\mathbf A} {\!\mathbf B}} \ge 
\widetilde{\mathbf V}_{\!\!{\mathbf M} {\!\mathbf M}},  
\end{equation}
which implies by (\ref{willianson}) that 
$ f(\tilde{\nu}^{-}_{\mathbf{M}}) \ge 
  f(\widetilde{ \mu }_{-})             \ge 
  f(\tilde{\nu}^{-}_{\mathbf{N}})      $ 
where $\widetilde{ \mu }_{-}$ is the symplectic 
eigenvalue defined in (\ref{simpeitp}) of 
$\widetilde{\mathbf V}_{\!\!{\mathbf A} {\!\mathbf B}}$ 
and $\tilde{\nu}^{-}_{\mathbf{M}}$ and $\tilde{\nu}^{-}_{\mathbf{N}}$
are the SE in (\ref{simsimpeitp}) of the symmetric states 
$\hat \rho_{  \mathbf M\!\mathbf M }$ and $\hat \rho_{  \mathbf N\!\mathbf N }$.
Obviously this works for the natural bounds (\ref{ord3}). 
It is interesting to note that even knowing that 
the EeoF is not the true EoF for GS \cite{PhysRevLett.101.220403}, 
it is bounded as if it were and this may be used to consider 
the EeoF as a good estimation for the true one. 
Numerical exploitations show that the estimation 
can be greater or smaller than the GeoF. 

Needless to say, the statement in (\ref{willianson}) 
can be used to sort and 
determine some bounds, {\it e.g.}, the ordering 
$ {\rm EoF} \left( \hat \rho_{   \mathbf B \!\mathbf B}  \right) \le 
  {\rm EoF} \left( \hat \sigma  \right) $ in Eq.~(\ref{ord2}) can be attained 
if one compares the SEs associated with the partially transposed states.
   
%
%

%

\section{Conclusions}
Starting with the convexity property of the EoF, 
we describe a simple method to construct lower and upper bounds to the 
Entanglement of Formation for general Gaussian states which has a clear 
physical interpretation in terms of the action of a noise channel. 
The same procedure is used to define what we called natural bounds 
since they are constructed using only the one-mode reduced CM of a two-mode 
GS. 
We have also demonstrated that the same bounds can be applied to the generalization of EoF, 
where it is considered a monotonically decreasing function of the smaller 
symplectic eigenvalue --- since we can not define it as a lower or an upper bound, 
we call it an estimation of the EoF (or the EeoF). 
For this we used the Williamson 
theorem for positive definite matrices, highlighting the underlying
mathematical character of the EeoF. 
We strongly believe that these results can be used in the direction to obtain 
a closed and analytical formula for the EoF of general (nonsymmetric) Gaussian 
states. This is currently under investigation. 

\begin{acknowledgments}
{This work is supported by the Brazilian funding agencies CNPq 
and FAPESP through the Instituto Nacional de Ci{\^e}ncia e 
Tecnologia - Informa\c{c}{\~a}o Qu{\^a}ntica (INCT-IQ).
F.N. wishes to acknowledge financial support from FAPESP (Proc. 2009/16369-8).}
The authors would like to thank G. Rigolin and M. F. Corn\'elio 
for insightful discussions. 
\end{acknowledgments}

\bibliographystyle{apsrev4-1}
\bibliography{bibliografia.bib}

\begin{thebibliography}{15}%
\makeatletter
\providecommand \@ifxundefined [1]{%
 \@ifx{#1\undefined}
}%
\providecommand \@ifnum [1]{%
 \ifnum #1\expandafter \@firstoftwo
 \else \expandafter \@secondoftwo
 \fi
}%
\providecommand \@ifx [1]{%
 \ifx #1\expandafter \@firstoftwo
 \else \expandafter \@secondoftwo
 \fi
}%
\providecommand \natexlab [1]{#1}%
\providecommand \enquote  [1]{``#1''}%
\providecommand \bibnamefont  [1]{#1}%
\providecommand \bibfnamefont [1]{#1}%
\providecommand \citenamefont [1]{#1}%
\providecommand \href@noop [0]{\@secondoftwo}%
\providecommand \href [0]{\begingroup \@sanitize@url \@href}%
\providecommand \@href[1]{\@@startlink{#1}\@@href}%
\providecommand \@@href[1]{\endgroup#1\@@endlink}%
\providecommand \@sanitize@url [0]{\catcode `\\12\catcode `\$12\catcode
  `\&12\catcode `\#12\catcode `\^12\catcode `\_12\catcode `\%12\relax}%
\providecommand \@@startlink[1]{}%
\providecommand \@@endlink[0]{}%
\providecommand \url  [0]{\begingroup\@sanitize@url \@url }%
\providecommand \@url [1]{\endgroup\@href {#1}{\urlprefix }}%
\providecommand \urlprefix  [0]{URL }%
\providecommand \Eprint [0]{\href }%
\providecommand \doibase [0]{http://dx.doi.org/}%
\providecommand \selectlanguage [0]{\@gobble}%
\providecommand \bibinfo  [0]{\@secondoftwo}%
\providecommand \bibfield  [0]{\@secondoftwo}%
\providecommand \translation [1]{[#1]}%
\providecommand \BibitemOpen [0]{}%
\providecommand \bibitemStop [0]{}%
\providecommand \bibitemNoStop [0]{.\EOS\space}%
\providecommand \EOS [0]{\spacefactor3000\relax}%
\providecommand \BibitemShut  [1]{\csname bibitem#1\endcsname}%
\let\auto@bib@innerbib\@empty
\bibitem [{\citenamefont {Horodecki}\ \emph {et~al.}(2009)\citenamefont
  {Horodecki}, \citenamefont {Horodecki}, \citenamefont {Horodecki},\ and\
  \citenamefont {Horodecki}}]{RevModPhys.81.865}%
  \BibitemOpen
  \bibfield  {author} {\bibinfo {author} {\bibfnamefont {R.}~\bibnamefont
  {Horodecki}}, \bibinfo {author} {\bibfnamefont {P.}~\bibnamefont
  {Horodecki}}, \bibinfo {author} {\bibfnamefont {M.}~\bibnamefont
  {Horodecki}}, \ and\ \bibinfo {author} {\bibfnamefont {K.}~\bibnamefont
  {Horodecki}},\ }\href {\doibase 10.1103/RevModPhys.81.865} {\bibfield
  {journal} {\bibinfo  {journal} {Rev. Mod. Phys.}\ }\textbf {\bibinfo {volume}
  {81}},\ \bibinfo {pages} {865} (\bibinfo {year} {2009})}\BibitemShut
  {NoStop}%
\bibitem [{\citenamefont {Horodecki}\ \emph {et~al.}(2000)\citenamefont
  {Horodecki}, \citenamefont {Horodecki},\ and\ \citenamefont
  {Horodecki}}]{PhysRevLett.84.2014}%
  \BibitemOpen
  \bibfield  {author} {\bibinfo {author} {\bibfnamefont {M.}~\bibnamefont
  {Horodecki}}, \bibinfo {author} {\bibfnamefont {P.}~\bibnamefont
  {Horodecki}}, \ and\ \bibinfo {author} {\bibfnamefont {R.}~\bibnamefont
  {Horodecki}},\ }\href {\doibase 10.1103/PhysRevLett.84.2014} {\bibfield
  {journal} {\bibinfo  {journal} {Phys. Rev. Lett.}\ }\textbf {\bibinfo
  {volume} {84}},\ \bibinfo {pages} {2014} (\bibinfo {year}
  {2000})}\BibitemShut {NoStop}%
\bibitem [{\citenamefont {Lloyd}\ and\ \citenamefont
  {Braunstein}(1999)}]{PhysRevLett.82.1784}%
  \BibitemOpen
  \bibfield  {author} {\bibinfo {author} {\bibfnamefont {S.}~\bibnamefont
  {Lloyd}}\ and\ \bibinfo {author} {\bibfnamefont {S.~L.}\ \bibnamefont
  {Braunstein}},\ }\href {\doibase 10.1103/PhysRevLett.82.1784} {\bibfield
  {journal} {\bibinfo  {journal} {Phys. Rev. Lett.}\ }\textbf {\bibinfo
  {volume} {82}},\ \bibinfo {pages} {1784} (\bibinfo {year}
  {1999})}\BibitemShut {NoStop}%
\bibitem [{\citenamefont {Giedke}\ \emph {et~al.}(2003)\citenamefont {Giedke},
  \citenamefont {Wolf}, \citenamefont {Kr\"uger}, \citenamefont {Werner},\ and\
  \citenamefont {Cirac}}]{PhysRevLett.91.107901}%
  \BibitemOpen
  \bibfield  {author} {\bibinfo {author} {\bibfnamefont {G.}~\bibnamefont
  {Giedke}}, \bibinfo {author} {\bibfnamefont {M.~M.}\ \bibnamefont {Wolf}},
  \bibinfo {author} {\bibfnamefont {O.}~\bibnamefont {Kr\"uger}}, \bibinfo
  {author} {\bibfnamefont {R.~F.}\ \bibnamefont {Werner}}, \ and\ \bibinfo
  {author} {\bibfnamefont {J.~I.}\ \bibnamefont {Cirac}},\ }\href {\doibase
  10.1103/PhysRevLett.91.107901} {\bibfield  {journal} {\bibinfo  {journal}
  {Phys. Rev. Lett.}\ }\textbf {\bibinfo {volume} {91}},\ \bibinfo {pages}
  {107901} (\bibinfo {year} {2003})}\BibitemShut {NoStop}%
\bibitem [{\citenamefont {Wolf}\ \emph {et~al.}(2004)\citenamefont {Wolf},
  \citenamefont {Giedke}, \citenamefont {Kr\"uger}, \citenamefont {Werner},\
  and\ \citenamefont {Cirac}}]{PhysRevA.69.052320}%
  \BibitemOpen
  \bibfield  {author} {\bibinfo {author} {\bibfnamefont {M.~M.}\ \bibnamefont
  {Wolf}}, \bibinfo {author} {\bibfnamefont {G.}~\bibnamefont {Giedke}},
  \bibinfo {author} {\bibfnamefont {O.}~\bibnamefont {Kr\"uger}}, \bibinfo
  {author} {\bibfnamefont {R.~F.}\ \bibnamefont {Werner}}, \ and\ \bibinfo
  {author} {\bibfnamefont {J.~I.}\ \bibnamefont {Cirac}},\ }\href {\doibase
  10.1103/PhysRevA.69.052320} {\bibfield  {journal} {\bibinfo  {journal} {Phys.
  Rev. A}\ }\textbf {\bibinfo {volume} {69}},\ \bibinfo {pages} {052320}
  (\bibinfo {year} {2004})}\BibitemShut {NoStop}%
\bibitem [{\citenamefont {Rigolin}\ and\ \citenamefont
  {Escobar}(2004)}]{PhysRevA.69.012307}%
  \BibitemOpen
  \bibfield  {author} {\bibinfo {author} {\bibfnamefont {G.}~\bibnamefont
  {Rigolin}}\ and\ \bibinfo {author} {\bibfnamefont {C.~O.}\ \bibnamefont
  {Escobar}},\ }\href {\doibase 10.1103/PhysRevA.69.012307} {\bibfield
  {journal} {\bibinfo  {journal} {Phys. Rev. A}\ }\textbf {\bibinfo {volume}
  {69}},\ \bibinfo {pages} {012307} (\bibinfo {year} {2004})}\BibitemShut
  {NoStop}%
\bibitem [{\citenamefont {Marian}\ and\ \citenamefont
  {Marian}(2008)}]{PhysRevLett.101.220403}%
  \BibitemOpen
  \bibfield  {author} {\bibinfo {author} {\bibfnamefont {P.}~\bibnamefont
  {Marian}}\ and\ \bibinfo {author} {\bibfnamefont {T.~A.}\ \bibnamefont
  {Marian}},\ }\href {\doibase 10.1103/PhysRevLett.101.220403} {\bibfield
  {journal} {\bibinfo  {journal} {Phys. Rev. Lett.}\ }\textbf {\bibinfo
  {volume} {101}},\ \bibinfo {pages} {220403} (\bibinfo {year}
  {2008})}\BibitemShut {NoStop}%
\bibitem [{\citenamefont {Eisert}\ and\ \citenamefont
  {Wolf}()}]{EisertWolf2005}%
  \BibitemOpen
  \bibfield  {author} {\bibinfo {author} {\bibfnamefont {J.}~\bibnamefont
  {Eisert}}\ and\ \bibinfo {author} {\bibfnamefont {M.}~\bibnamefont {Wolf}},\
  }\href@noop {} {}\bibinfo {note} {ArXiv:quant-ph/0505151 (2005)}\BibitemShut
  {NoStop}%
\bibitem [{\citenamefont {Simon}(2000)}]{PhysRevLett.84.2726}%
  \BibitemOpen
  \bibfield  {author} {\bibinfo {author} {\bibfnamefont {R.}~\bibnamefont
  {Simon}},\ }\href {\doibase 10.1103/PhysRevLett.84.2726} {\bibfield
  {journal} {\bibinfo  {journal} {Phys. Rev. Lett.}\ }\textbf {\bibinfo
  {volume} {84}},\ \bibinfo {pages} {2726} (\bibinfo {year}
  {2000})}\BibitemShut {NoStop}%
\bibitem [{\citenamefont {Serafini}\ \emph {et~al.}(2004)\citenamefont
  {Serafini}, \citenamefont {Illuminati},\ and\ \citenamefont
  {Siena}}]{0953-4075-37-2-L02}%
  \BibitemOpen
  \bibfield  {author} {\bibinfo {author} {\bibfnamefont {A.}~\bibnamefont
  {Serafini}}, \bibinfo {author} {\bibfnamefont {F.}~\bibnamefont
  {Illuminati}}, \ and\ \bibinfo {author} {\bibfnamefont {S.~D.}\ \bibnamefont
  {Siena}},\ }\href {http://stacks.iop.org/0953-4075/37/i=2/a=L02} {\bibfield
  {journal} {\bibinfo  {journal} {Journal of Physics B: Atomic, Molecular and
  Optical Physics}\ }\textbf {\bibinfo {volume} {37}},\ \bibinfo {pages} {L21}
  (\bibinfo {year} {2004})}\BibitemShut {NoStop}%
\bibitem [{\citenamefont {Adesso}\ and\ \citenamefont
  {Illuminati}(2007)}]{1751-8121-40-28-S01}%
  \BibitemOpen
  \bibfield  {author} {\bibinfo {author} {\bibfnamefont {G.}~\bibnamefont
  {Adesso}}\ and\ \bibinfo {author} {\bibfnamefont {F.}~\bibnamefont
  {Illuminati}},\ }\href {http://stacks.iop.org/1751-8121/40/i=28/a=S01}
  {\bibfield  {journal} {\bibinfo  {journal} {Journal of Physics A:
  Mathematical and Theoretical}\ }\textbf {\bibinfo {volume} {40}},\ \bibinfo
  {pages} {7821} (\bibinfo {year} {2007})}\BibitemShut {NoStop}%
\bibitem [{\citenamefont {Caruso}\ \emph {et~al.}(2008)\citenamefont {Caruso},
  \citenamefont {Eisert}, \citenamefont {Giovannetti},\ and\ \citenamefont
  {Holevo}}]{1367-2630-10-8-083030}%
  \BibitemOpen
  \bibfield  {author} {\bibinfo {author} {\bibfnamefont {F.}~\bibnamefont
  {Caruso}}, \bibinfo {author} {\bibfnamefont {J.}~\bibnamefont {Eisert}},
  \bibinfo {author} {\bibfnamefont {V.}~\bibnamefont {Giovannetti}}, \ and\
  \bibinfo {author} {\bibfnamefont {A.~S.}\ \bibnamefont {Holevo}},\ }\href
  {http://stacks.iop.org/1367-2630/10/i=8/a=083030} {\bibfield  {journal}
  {\bibinfo  {journal} {New Journal of Physics}\ }\textbf {\bibinfo {volume}
  {10}},\ \bibinfo {pages} {083030} (\bibinfo {year} {2008})}\BibitemShut
  {NoStop}%
\bibitem [{\citenamefont {Caves}\ and\ \citenamefont
  {Wodkiewicz}()}]{CavesWodf2004}%
  \BibitemOpen
  \bibfield  {author} {\bibinfo {author} {\bibfnamefont {C.}~\bibnamefont
  {Caves}}\ and\ \bibinfo {author} {\bibfnamefont {K.}~\bibnamefont
  {Wodkiewicz}},\ }\href@noop {} {}\bibinfo {note} {ArXiv:quant-ph/0409063
  (2004)}\BibitemShut {NoStop}%
\bibitem [{\citenamefont {de~Almeida}(1998)}]{deAlmeida1998265}%
  \BibitemOpen
  \bibfield  {author} {\bibinfo {author} {\bibfnamefont {A.~M.}\ \bibnamefont
  {de~Almeida}},\ }\href {\doibase 10.1016/S0370-1573(97)00070-7} {\bibfield
  {journal} {\bibinfo  {journal} {Physics Reports}\ }\textbf {\bibinfo {volume}
  {295}},\ \bibinfo {pages} {265 } (\bibinfo {year} {1998})}\BibitemShut
  {NoStop}%
\bibitem [{\citenamefont {de~Gosson}\ and\ \citenamefont
  {Luef}(2009)}]{deGosson2009131}%
  \BibitemOpen
  \bibfield  {author} {\bibinfo {author} {\bibfnamefont {M.}~\bibnamefont
  {de~Gosson}}\ and\ \bibinfo {author} {\bibfnamefont {F.}~\bibnamefont
  {Luef}},\ }\href {\doibase 10.1016/j.physrep.2009.08.001} {\bibfield
  {journal} {\bibinfo  {journal} {Physics Reports}\ }\textbf {\bibinfo {volume}
  {484}},\ \bibinfo {pages} {131 } (\bibinfo {year} {2009})}\BibitemShut
  {NoStop}%
\end{thebibliography}%
\end{document}